\documentclass[twocolumn, pra]{revtex4-1}
\usepackage{graphicx}
\usepackage{epsfig}
\usepackage{color}

\begin{document}
\title{Effective spin-chain model for strongly interacting one-dimensional atomic gases with an arbitrary spin}
%\title{Strongly Interacting High-Spin Atomic Gases in One-Dimensional Traps\\ Effective Spin-Chain Model and the Effect of Spin-Orbit Coupling}
\author{Lijun Yang and Xiaoling Cui}
\email{xlcui@iphy.ac.cn}
\affiliation{Beijing National Laboratory for Condensed Matter Physics, Institute of Physics, Chinese Academy of Sciences, Beijing, 100190, People's Republic of China}
\date{\today}

\begin{abstract}
We present a general form of the effective spin-chain model for strongly interacting atomic gases with an arbitrary spin in the one-dimensional(1D) traps. In particular, for high-spin systems the atoms can collide in multiple scattering channels, and we find that the resulted form of spin-chain model generically follows the same structure as that of the interaction potentials.  This is a unified form working for any spin, statistics (Bose or Fermi) and confinement potentials. We adopt the spin-chain model to reveal both the ferromagnetic(FM) and anti-ferromagnetic(AFM) magnetic orders for strongly interacting spin-1 bosons in 1D traps. We further show that by adding the spin-orbit coupling, the FM/AFM orders can be gradually destroyed and eventually the ground state exhibits universal spin structure and contacts that are independent of the strength of spin-orbit coupling.

 \end{abstract}

\maketitle

\section{Introduction}

Strongly interacting system is known to be notoriously difficult to
solve in physics. There is one exception, however, for the infinite
coupling of one-dimensional (1D) system that the exact
solutions of bosons and fermions can be constructed taking advantage of the peculiar feature of femionalization\cite{Girardeau1,Girardeau2, Sengstock, Chen, Girardeau3}.
%Bose-Fermi mapping or Fermi-Fermi mapping\cite{Girardeau1,2}.
%The construction has taken advantages of the impenetrable feature (i.e., the fermionlization) of particles in the presence of the hard-core repulsion, i.e., the so-called Tonks-Girardeau limit.
In cold atoms experiments, strongly interacting Tonks-Girardeau gases have been realized in both spinless bosons\cite{Bloch, Weiss, Haller} and spin-1/2 fermions\cite{Jochim1,Jochim2, Jochim3}. A fascinating property with an infinite coupling (hard-core interaction) is that,
the ground states of a spinful system are highly degenerate and their wave functions share the form of
\begin{equation}
\Psi(x_1,\mu_1;...;x_N,\mu_N)=\phi_F(x_1,...,x_N)\psi(x_1,\mu_1;...;x_N,\mu_N).
\end{equation}
Here $x_i$ and $\mu_i$ are position and spin of the $i$-th particle $(i=1,...,N)$; $\phi_F$ is the Slater determinant made up of the lowest N-level of eigenstates in 1D system, a common factor of all degenerate wave functions $\Psi$. While the energy of $\Psi$ is solely given by $\phi_F$, the $\psi$ part uniquely describes the distribution of spins in the coordinate space and determines the degeneracy of the system.
%Interestingly, due to the large degeneracy at infinite coupling, the spin-1/2 fermions are found to undergo the ferromagnetic transition as tuning the coupling strength across this point\cite{Cui1, Cui2}.
The large degeneracy has been shown to facilitate the Ferromagnetic transition of spin-1/2 fermions as tuning the coupling strength across this critical point\cite{Cui1, Cui2}.

%Though the physics right at the 1D infinite coupling is well understood, much less is known when the interaction is slightly away from this special point.
Apart from the infinite coupling case, it is interesting and also more practical to learn about the physics in the regime of large but finite couplings. This regime is more realistic to achieve in experiments, which can be adiabatically connected to the non-interacting limit as tuning the interaction strength. With finite couplings, the magnetic properties of bosons and fermions are very different. General theorems have shown that the ground state for spin-1/2 fermions is with the lowest total spin\cite{Lieb-Mattis}, while for iso-spin 1/2 bosons is ferromagnetic\cite{FM1,FM2,FM3}. Accordingly, the effective anti-ferromagnetic(AFM) and ferromagnetic(FM) spin-spin exchange interactions have been successfully extracted from the Bethe-ansatz solutions of strongly coupling spin-1/2 fermions\cite{Guan_f} and bosons\cite{Guan_b}. Recently, taking advantage of the high controllability of a few particles in the trapped ultra-cold systems\cite{Jochim_expe}, a number of studies have revealed the energy spectra and correlation effects for a few spin-1/2 fermions by numerical simulations\cite{Blume, Conduit, Lewenstein, Zinner1}. In the strong coupling regime, an effective Heisenberg spin-chain model has also been deduced\cite{Zinner2, Santos, Pu, Levinsen}. The resulted anti-ferromagnetic correlation has recently been confirmed through the tunneling measurement of a few spin-1/2 fermions in 1D traps\cite{Jochim_chain}. Moreover, the effective models for spin-1/2 bosons\cite{Zinner3, Levinsen2} and for higher-spin case with SU(N) symmetry\cite{Zinner2, Santos} have also been discussed.

In this work, we will present a general form of the effective spin-chain model for strongly interacting trapped atomic gases with an arbitrary spin and arbitrary statistics (Bose or Fermi).
%bosons/fermions with an arbitrary spin.
In particular, for high-spin atomic systems, the multi-channel interactions can break the SU(N) symmetry and the ground state can have a wide variety of magnetic orders, including FM, AFM or even intriguing ones, depending on the relative coupling strength between different scattering channels.
We find that the resulted form of effective spin-chain model generically follows the same structure as that of the interaction potentials classified by scattering channels, thus respects the symmetry of original Hamiltonian for trapped systems. This form can be applied to any spin-value, Bose or Fermi statistics and confinement potentials.
%Our derivation reveals intrinsic relation between the Bose/Fermi statistics, scattering channels and the resulted form of the model. This also allows us to present a general form of the spin-chain model for an arbitrarily high-spin system.
%spin-1/2 fermions, our result is consistent with previous ones giving the anti-ferromagnetic Heisenberg chain model\cite{ Zinner2, Santos, Pu, Levinsen}, which has recently be confirmed in the tunneling experiment of a few particles in 1D traps\cite{Jochim_chain}. For two-component bosons with spin-independent interactions, we find a ferromagnetic Heisenberg chain model. They are respectively consistent with the two exact theorems saying that the ground state of fermions and bosons are with the smallest\cite{Lieb-Mattis} and largest total spin\cite{FM1,FM2,FM3}. For high-spin systems, we find the model generically follow the similar form with that for the Mott phase in optical lattices, giving more physical insight to the essense of the spin-chain model.
%Compared to spin-1/2 fermions, the high-spin allows particles scattering in a multiple number of total spin channels and thus the ground state has a wide variety of spin-spin correlations, which can be ferromagnetic, anti-ferromagnetic or even more intriguing ones, depending on the relative coupling strength between different scattering channels. In this work,
Take the strongly interacting spin-1 bosons for example, we adopt the effective spin-chain model and reveal both FM and AFM correlations by choosing different interaction strengths for different cold atoms.
%, from the both the energy spectra and the spin correlation for FM and AFM  show how the FM or AFM ground state can be obtained by tuning  energy spectra and the spin texture/correlation vary as changing the scattering strengths in different channels.
%For instance, for spin-1 bosons, there are two channels:..., can be both FM and AFM, given the strength of $S=0$ channel scattering larger or smaller than $S=2$ channel scattering.

Based on the spin-chain model, we further study the effect of spin-orbit coupling(SOC), which has been realized in cold atoms experiments by the two-photon Raman processes\cite{Ian_1, Ian_2, Shuai, Jing, Martin,Engels}. For spin-1 bosons, we will show how the FM or AFM correlations being destroyed as increasing the SOC strength. In the SOC-dominated regime, we obtain universal spin texture and contacts, which are independent of the actual SOC strength. These results are consistent with previous studies on the spin-orbit coupled spin-1/2 fermions at infinite coupling \cite{Cui3, Blume2}.

%Previous studies have shown that an infinitesimal soc will be able to induce giant spin-texture for spin-1/2 fermions with infinite repulsion. The spin texture is universal but not depending on the strength of doc strength.  It is thus interesting to study the interplay of soc with spin-chain model for large but finite interactions, and how the universal texture, if any, appears for high-spin when gradually increasing the soc strength.

The rest of the paper is organized as follows. In section II, we present the derivation of the effective spin-chain model with an arbitrary spin. We will first take the spin-1/2 (two-component) fermions and bosons as a starting point, to get insight to the general structure of spin-chain models with an arbitrary spin. In section III, based on the effective spin-chain model, we study the system of spin-1 bosons and show the ground state exhibiting FM or AFM correlations. We further study the interplay effect of strong interaction and SOC to the ground state properties of spin-1 bosons under the effective spin-chain model.
%and investigate the ground state properties as a function of increasing SOC strength, and show that the universal regime can be obtained as tuning SOC to be larger than the coupling in the spin-chain model.
Finally we summarize in section IV.

\section{Effective spin-chain model}

%We first write down the general Hamilnotian for spin-$f$ particles in the 1D trap as $H=H_0+U$, with
%\begin{eqnarray}
%H_0&=&\sum_i \left( -\frac{\hbar^2}{2m} \frac{\partial^2}{\partial x_i^2} + V_T(x_i)  \right), \label{H0}\\
%U&=&\sum_{S=0}^{2f}\sum_{M=-S}^{S} g_{SM}  \sum_{i<j} \delta(x_i-x_j) P_{SM}(i,j). \label{U}
%\end{eqnarray}
%Here $V_T$ is the trapping potential; $g_{SM}$ is the coupling constant for two particles scattering with total spin $S$ and total magnetization $M$; $P_{SM}(i,j)$ is the projection operator for particles $i,j$ scattering in the $\{SM\}$ sector. In this work we consider the strong couplings in all $\{SM\}$ sectors, which can be achieved through the confinement induced resonance\cite{Olshanii} or the preparation of very dilute gas.

We first write down the non-interacting Hamiltonian for atoms confined in the 1D trap,
\begin{eqnarray}
H_0&=&\sum_i \left( -\frac{\hbar^2}{2m} \frac{\partial^2}{\partial x_i^2} + V_T(x_i)  \right), \label{H0}
%U&=&\sum_{S=0}^{2f}\sum_{M=-S}^{S} g_{SM}  \sum_{i<j} \delta(x_i-x_j) P_{SM}(i,j). \label{U}
\end{eqnarray}
Here $V_T$ is a spin-independent trapping potential and in this paper we consider a harmonic trap with trapping frequency $ \omega_T$ and characteristic length $a_{ho}=1/\sqrt{m\omega_T}$. The interaction is generally characterized by the coupling constant $g$, and for the high-spin atoms there can be multiple scattering channels with multiple coupling constants. In this work we consider the large repulsive coupling constants in all scattering channels, which can be achieved through the confinement induced resonance\cite{Olshanii} or the preparation of very dilute gas.

The basic idea for the construction of an effective spin-chain model is that, the physics in the vicinity of infinite coupling $g\rightarrow \infty$ can be well deduced from the known $g=\infty$ limit, by treating $1/g$ as a small parameter in the framework of the perturbation theory. In this way the wave function to the zeroth order of $1/g$ can be approximated as certain superposition of the degenerate states at $1/g=0$, which leads to an energy functional up to the linear $1/g$ and gives the effective model. This idea has been successfully applied to the spin-1/2 fermion case and leads to the AFM Heisenberg spin-chain model therein\cite{Zinner2, Santos, Pu, Levinsen}.

In this section, we will first give a detailed introduction to the degenerate ground states at infinite coupling, as they are essential for the construction of spin-chain model when slightly away from this special point. These degenerate states are classified by the order of spins in coordinate space and thus named as the spin-ordered states as in Ref.\cite{Cui3}. Using these states, we will re-derive the spin-chain model for the spin-1/2 fermions and bosons as a starting point. Finally, we will extend our derivation to an arbitrary spin case and present a general form of spin-chain model for both fermions and bosons.

\subsection{Spin-ordered state}

To conveniently enumerate the degenerate ground states at infinite coupling of 1D systems, we define the spin-ordered state:
\begin{equation}
|\{\xi_1, \xi_2, \cdots ,\xi_N\} \rangle\equiv |\vec{\xi}\rangle, \label{spin-order_1}
\end{equation}
In this state,  a sequence  of spins $\xi_{1}$, $\xi_{2},\cdots,\xi_{N}$ is placed in order in the 1D coordinate space. Explicitly, its wave function is written as
\begin{eqnarray}
&&\langle x_1,\cdots,x_N; \mu_1,\cdots,\mu_N | \vec{\xi} \rangle \nonumber\\
&=&\sum_{P}
 \theta(x_{P_1}, x_{P2}, \cdots, x_{P_N})
\prod_i
 \delta_{\xi_i,\mu_{P_i}},  \label{spin-order}
\end{eqnarray}
 where $P$ is a permutation of the integers $(1,2, \cdots, N)$, and
\begin{eqnarray}
\theta(x_{P_1}, x_{P2}, \cdots, x_{P_N}) & =&1 \,\,\,\,\,  {\rm if}  \,\,\,\,\,  x_{P_1}< x_{P2}<\cdots<x_{P_N}, \nonumber  \\
   &=& 0 \,\, \,\,\,\,  {\rm otherwise}.  \hspace{0.7in}
\end{eqnarray}
%By considering different spin orders in Eq.\ref{spin-order_1}, one can cover all the degenerate ground states at infinite coupling.
%Moreover, as the particles are inpenetrable with each other, different spin orders decouple with each other, and thus these states form a complete and orthogonal set of eigenstates for the lowest energy space.

The spin-ordered state as defined in (\ref{spin-order_1}) is symmetric under the simultaneous change of the coordinate and spin of any two particles. It is then straightforward to construct the degenerate ground states of bosons and fermions at infinite coupling using this class of states:
\begin{eqnarray}
\Psi_B^{\xi}&=|\phi_F(x_1, x_2, \cdots, x_N)| \langle x_1,\cdots,x_N; \mu_1,\cdots,\mu_N |\vec{\xi} \rangle, \label{Psi_b}\\
\Psi_F^{\xi}&=\phi_F(x_1, x_2, \cdots, x_N)  \langle  x_1,\cdots,x_N; \mu_1,\cdots,\mu_N |\vec{\xi}\rangle, \label{Psi_f}
\end{eqnarray}
here $\Psi_B$($\Psi_F$) is the wave function of bosons (fermions) obeying Bose (Fermi) statistics;  $\phi_F$ is the Slater determinant composed by the lowest N-level of eigenstates of $H_0$ in Eq.\ref{H0}:
\begin{eqnarray}
\phi_F(x_1, x_2, \cdots, x_N)&=&\frac{1}{\sqrt{N!}}D(x_1, x_2, \cdots, x_N) \nonumber\\
&=& \prod_{i<j} (x_i-x_j) F(x_1, x_2, \cdots, x_N),
\label{phi_F}
\end{eqnarray}
with $F(\{x_i\})$ a fully symmetric function (with respect to the exchange of any two coordinates).
In Eqs.(\ref{Psi_b},\ref{Psi_f}), the $\phi_F$ part determines the ground state energy of the system, while the $\xi$ part uniquely determines the spin distribution/order of system. By considering different spin orders in $\xi$, one can cover all the degenerate ground states at infinite coupling. %At infinite coupling, the spin-ordered states are the eigen-states of the system, with the energy uniquely determined by the $\phi_F$ part. Thus, the states with different spin-order have the same energy and they are all degenerate. By enumerating all spin-orders, one can cover all degenerate ground states.
Moreover, it is easy to check that the states with different spin orders are orthogonal to each other.
%$\langle \vec{\xi}'| \vec{\xi}\rangle= \prod_{i=1}^{N} \delta_{\xi_{i}', \xi_{i}}$.
Therefore, the wave functions as Eqs.(\ref{Psi_b},\ref{Psi_f}) constitute a complete and orthogonal set of basis in ground state manifold of particles with infinite coupling strength.

Given Eqs.(\ref{Psi_b},\ref{Psi_f}),
%spin-ordered state (\ref{spin-order-1}),
one can compute the particle density at the $i-$th spin order:
\begin{eqnarray}
n_{i}(x) &= \int {\rm d}{\bf  x} |D|^2 \theta(x_1,x_2, \cdots,x_i,\cdots,x_N) \delta(x-x_{i}), \,\,\,\, \,\,    \label{ni}
\end{eqnarray}
with $d{\bf x}=\prod_{i=1}^N dx_i$. $n_i(x)$ gives the probability of finding the $i-$th ordered particle at position $x$ and we have $ \int dx n_i(x)=1$. In Fig.1, we show $n_{i}(x)$ (with $i=1,2,\cdots,N$) for $N=6$ particles in a 1D harmonic trap. It is found that the density peak for each spin order is well separated from each other, and $n_i(x)$ can be well approximated by a Gaussian function\cite{Cui3}:
\begin{equation}
n_i(x)\rightarrow \frac{1}{\sqrt{\pi}\sigma_i} e^{-(x-{\bar x}_i)^2/\sigma_i^2},  \label{ni}
\end{equation}
here ${\bar x}_i=\int x n_i(x) dx$ is the averaged location of the $i-$th particle, and $\sigma_i$ is the width of the density distribution for the $i-$th particle which can be obtained by fitting $n_i(x)$ with above function. In Table I, we present the result of $x_i$, $\bar{x}_{i+1}-\bar{x}_i$ and $\sigma_i$ for $N=4,6,8$ particles. It is found that the neighboring $\bar{x}_i$ and $\bar{x}_{i+1}$ are nearly equally spaced by a distance $d$, and each $\sigma_i$ is typically of order of $d$ and varies little for different $i$.

\begin{figure}[t]
\includegraphics[height=7cm,width=8.5cm]{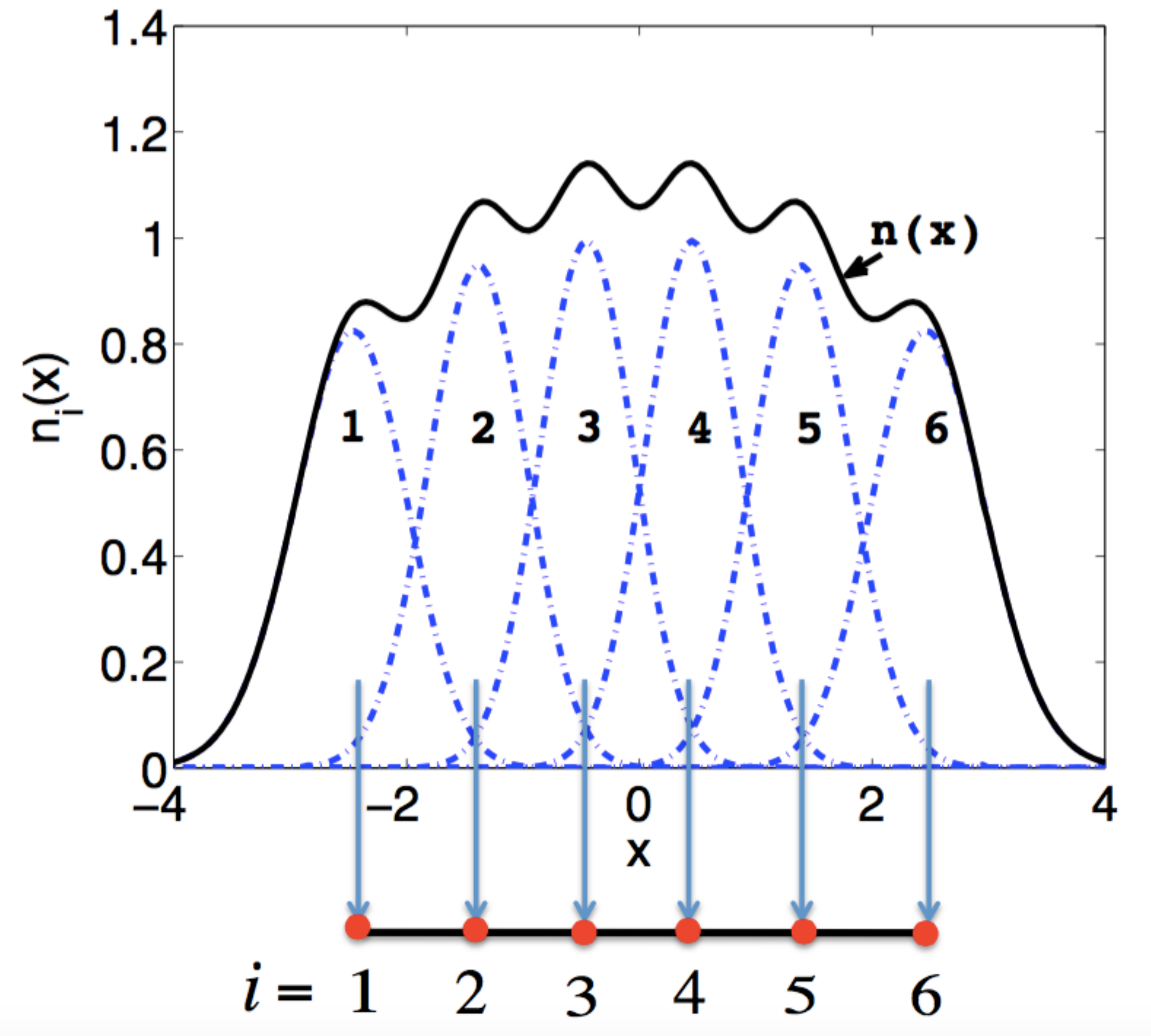}
\caption{(Color online). The particle density $n_i(x)$ (blue dashed lines) for $N=6$ particles in a harmonic trap. $x$ and $n_i$ are respectively in the unit of $a_{ho}$ and $1/a_{ho}$, with $a_{ho}$ the confinement length of the trapping potential. The black curve shows the total density $n(x)\equiv \sum_i n_i(x)$. Here the spin-order index $i$ is mapped to the site index $i$ in the effective spin-chain model.  } \label{fig1}
\end{figure}

\begin{table}[t]%[!hbp]
\begin{tabular}{|c|c|c|c|c|}
\hline
\multicolumn{5}{|c|}{$N=4$}\\
  \hline
   $i$ & 1& 2&3&4\\
  \hline
  $\bar{x_i}$ & -1.751 & -0.551 & 0.551& 1.751 \\
  \hline
  $\bar{x}_{i+1}-\bar{x_i}$  & 1.200 & 1.102 & 1.200&-\\
  \hline
  $\sigma_i$ & 0.739 & 0.653 & 0.653 & 0.739 \\
  \hline
\end{tabular}\\
\begin{tabular}{|c|c|c|c|c|c|c|}
\hline
\multicolumn{7}{|c|}{$N=6$}\\
  \hline
     $i$ & 1& 2&3&4&5&6\\
  \hline
  $\bar{x_i}$ & -2.451 & -1.466& -0.450& 0.450 &1.466& 2.451 \\
  \hline
  $\bar{x}_{i+1}-\bar{x_i}$ & 0.985 & 1.016 & 0.900 &  1.016 & 0.985& -\\
  \hline
  $\sigma_i$ & 0.684 & 0.595 & 0.567 & 0.567 & 0.595 & 0.684 \\
  \hline
\end{tabular}\\
\begin{tabular}{|c|c|c|c|c|c|c|c|c|}
\hline
\multicolumn{9}{|c|}{$N=8$}\\
  \hline
     $i$& 1& 2&3&4&5&6&7&8\\
  \hline
  $\bar{x_i}$ & -3.049 & -2.044 & -1.189& -0.394  & 0.394 &1.189 &2.044 & 3.049\\
    \hline
  $\bar{x}_{i+1}-\bar{x_i}$ & 1.005 & 0.855 & 0.795 &  0.780 & 0.795 & 0.855 & 1.005& -\\
  \hline
  $\sigma_i$ & 0.648 & 0.556 & 0.522 & 0.508 &0.508 & 0.522 & 0.556 & 0.648 \\
  \hline
\end{tabular}
\caption{The averaged location $\bar x_i$, the difference $\bar{x}_{i+1}-\bar{x_i}$ and variance $\sigma_i$ for $N=4,6,8$.}
\end{table}

When limited to the ground state manifold, we can map the spin-order index $i$ to the site index $i$, and thus the physics with spin-ordered state can be mapped to that under an effective spin-chain model. Here we like to point out an essential difference between the effective spin-chain and the real lattice configuration, that the particle density at each "site" of the effective spin-chain can actually spread over the inter-particle spacing of atoms in the harmon trap\cite{note_density}, while in the real lattice case just localize around each lattice site. The large spreading in the effective spin-chain case can lead to strong interference of spins between neighboring orders and give rise to exotic spin density profiles \cite{Cui3}.

\subsection{Effective spin chain model for spin-1/2}

In this section, we will derive the spin-chain model for spin-1/2 system, which was previously studied in Ref.\cite{Zinner2, Santos, Pu, Levinsen, Zinner3, Levinsen2}. Here we will adopt a systematic method in order for easy generalization to the high-spin case.

For spin-1/2 system with total particle number $N=N_{\uparrow}+N_{\downarrow}$, the number of different spin-ordered states, i.e., the ground state degeneracy  at infinite coupling, is $N_{\rm dg}=N!/(N_{\uparrow}! N_{\downarrow}!)$.

\subsubsection{spin-1/2 fermions}

Due to Fermi statistics, the contact interaction for spin-1/2 fermions only occurs in the spin-singlet channel with coupling constant $g$. In fact, in this case one can directly write the interaction as
\begin{equation}
U=g\sum_{i<j} \delta(x_i-x_j),
\end{equation}
because the particles scattering in other channels (other than spin-singlet) will be automatically ruled out by the asymmetric feature of the wave function (or the Fermi statistics).

For large $g$ (and thus small $1/g$), the many-body wave function can be written as certain superposition of the degenerate ground states (Eq.\ref{Psi_f}) at $1/g=0$:
\begin{eqnarray}
\Psi(\{x_i\};\{\mu_i\})=\phi_F(\{x_i\}) \sum_{k=1}^{N_{\rm dg}} a_k \langle \{x_i\};\{\mu_i\} |\vec{\xi}_k \rangle  .
\label{psi_f_many}
\end{eqnarray}
At small $1/g$, the associated energy can be expanded as:
\begin{equation}
E=E_0-\frac{\kappa}{g}, \label{energy}
\end{equation}
where $E_0$ is the degenerate energy at $1/g=0$ (determined by $\phi_F$), and $\kappa$ is proportional to the Tan's contact in 1D\cite{Contact_1D}:
\begin{equation}
\kappa= \frac{\partial E}{\partial (-1/g)}=g^2 \frac{\partial E}{\partial g}.  \label{kappa}
\end{equation}
In the following, we will aim at expressing $\kappa$ in terms of the coefficients $\{a_k\}$ in Eq.(\ref{psi_f_many}), which is essential to the construction of effective spin-chain model.

Applying the Hellmann-Feynman theorem \cite{HF} to Eq.(\ref{kappa}), we obtain
\begin{eqnarray}
\kappa&=&\lim_{g \rightarrow \infty} g^2  \int d{\bf x} \sum_{i<j}\delta(x_i-x_j) |\Psi(\{x_i\};\{\mu_i\})|^2   \nonumber\\
&=&  \frac{N(N-1)}{2}   \lim_{g \rightarrow \infty} g^2  \int d{\bf x} \delta(x_i-x_j) |\Psi(\{x_i\};\{\mu_i\})|^2   \label{kappa2}.\nonumber\\
\end{eqnarray}
%here $d{\bf x}=dx_1dx_2...dx_N$.
The integral in above equation can be obtained by considering the Schrodinger equation:
\begin{eqnarray}
&&\left( -\sum_{k\neq i,j} \frac{\hbar^2}{2m} \frac{\partial^2}{\partial x_k^2} +g\sum_{(k,l)\neq(i,j)} \delta(x_k-x_l) +\sum_k V_T(x_k)\right. \nonumber\\
&&\left.  \ \ - \frac{\hbar^2}{4m} \frac{\partial^2}{\partial X_{ij}^2} -\frac{\hbar^2}{m} \frac{\partial^2}{\partial x_{ij}^2} + g\delta(x_{ij}) \right) \Psi(\{x_i\};\{\mu_i\})=0,\nonumber\\
\end{eqnarray}
with $x_{ij}=x_i-x_j, X_{ij}=(x_i+x_j)/2$ respectively represent the relative and center-of-mass motions of $i$ and $j$ particles. The boundary condition around $x_{ij}=0$ gives  %we have
\begin{equation}
\frac{\hbar^2}{m} \Big( \frac{\partial \Psi}{\partial x_{ij}}|_{x_{ij}=0^+}-\frac{\partial \Psi}{\partial x_{ij}}|_{x_{ij}=0^-}\Big) =g \Psi |_{x_{ij}=0}.
\end{equation}
So we get $\kappa$ in Eq.(\ref{kappa2}) as
\begin{eqnarray}
\kappa=\frac{N(N-1)}{2} (\frac{\hbar^2}{m})^2 \int d{\bf x}  \Big|\frac{\partial \Psi}{\partial x_{ij}}|_{x_{ij}=0^-}^{x_{ij}=0^+} \Big|^2 .        \label{kappa_many}
\end{eqnarray}
Given the property of $\phi_F$ in Eq.(\ref{phi_F}), $\kappa$ can be further reduced to:
\begin{eqnarray}
&&\kappa=\frac{N(N-1)}{2} (\frac{\hbar^2}{m})^2 \int d{\bf x}  \Big|\frac{\partial \phi_F}{\partial x_{ij}}|_{x_{ij}=0} \Big|^2 \nonumber\\
&&\Big|\sum_{k=1}^{N_{\rm dg}} a_k \Big( \langle \{x_i\};\{\mu_i\} |\vec{\xi}_k \rangle|_{x_{ij}=0^+}-\langle \{x_i\};\{\mu_i\} |\vec{\xi}_k \rangle|_{x_{ij}=0^-} \Big) \Big|^2 .        \label{kappa_many_2}\nonumber\\
\end{eqnarray}
Due to the inclusion of $\frac{\partial \phi_F}{\partial x_{ij}}$, it is easy to check that only when the two coordinates, $x_i$ and $x_j$, stay in the neighboring order in the wave function, can they have contribution to $\kappa$. Assume $x_i$ and $x_j$ stay in the $l-$th and $(l+1)-$th order in $\Psi$, we denote their contribution to $\kappa$ as $\kappa_l$. For two particles there are four  spin-ordered states:
\begin{equation}
|\{ \uparrow \uparrow \}\rangle,\ \ |\{ \uparrow \downarrow \}\rangle,\ \ |\{ \downarrow \uparrow \}\rangle,\ \ |\{ \downarrow\downarrow \}\rangle;
\end{equation}
and they can form one singlet and three triplets:
\begin{eqnarray}
|00\rangle_{l,l+1}&=& \frac{| \{ \uparrow \downarrow \}\rangle -| \{ \downarrow\uparrow \}\rangle}{\sqrt{2}}; \nonumber\\
|11\rangle_{l,l+1}&=&| \{ \uparrow \uparrow \}\rangle ; \nonumber\\
|10\rangle_{l,l+1}&=& \frac{| \{ \uparrow\downarrow\}\rangle +| \{ \downarrow\uparrow \}\rangle}{\sqrt{2}}; \nonumber\\
|1,-1\rangle_{l,l+1}&=&| \{ \downarrow \downarrow \}\rangle . \nonumber
\end{eqnarray}
According to the definition of spin-ordered states in Eq.(\ref{spin-order}), above states can be simplified as:
\begin{eqnarray}
|00\rangle_{l,l+1}&\rightarrow& \frac{| \uparrow_i\downarrow_j \rangle -| \downarrow_i\uparrow_j\rangle}{\sqrt{2}} \Big( \theta(x_i,x_j)-\theta(x_j,x_i) \Big); \nonumber\\
|11\rangle_{l,l+1}&\rightarrow&| \uparrow_i \uparrow_j \rangle ; \nonumber\\
|10\rangle_{l,l+1}&\rightarrow& \frac{| \uparrow_i \downarrow_j \rangle +|  \downarrow_i \uparrow_j \rangle}{\sqrt{2}}; \nonumber\\
|1,-1\rangle_{l,l+1}&\rightarrow&|  \downarrow_i \downarrow_j \rangle . \nonumber
\end{eqnarray}
It is then easy to see that only the singlet state, $|00\rangle$, can contribute to $\kappa_l$ in Eq.(\ref{kappa_many_2}).  Physically, this is due to the Fermi statistics and the asymmetric feature of the fermionic wave function (\ref{psi_f_many}).

As the spin-ordered states in the wave function (\ref{psi_f_many}) can be classified according to the total spin and total magnetization of the $l$-th and $(l+1)$-th ordered particles, we can write
\begin{eqnarray}
\sum_{k} a_k |\vec{\xi}_k \rangle  \rightarrow \sum_{n} a_n^{SM} |SM \rangle_{l,l+1} |\vec{\xi}'_n \rangle
\label{psi_f_many_3}
\end{eqnarray}
here $|SM\rangle$ can be $|00\rangle,\ |11\rangle, |10\rangle,\ |1,-1\rangle$; $\vec{\xi}'$ means the spin-ordered states for the other order numbers except $l$ and $l+1$. Based on (\ref{psi_f_many_3}), we can obtain $\kappa_l$ as:
\begin{eqnarray}
\kappa_l&=&\frac{N!}{2} (\frac{\hbar^2}{m})^2 \int d{\bf x}  \Big|\frac{\partial \phi_F}{\partial x_{ij}}|_{x_{ij}=0} \Big|^2 \theta(...<x_i=x_j<...)\nonumber\\
&& \ \ \ \ \ \ \ \ \sum_n\Big|2a_n^{00}\Big|^2,        \label{kappa_l}
\end{eqnarray}
here $x_i$ is at the $l$-th order in the $\theta$-function (i.e., there are $(l-1)$ number of particles with coordinates smaller than $x_i$). The contribution of these two order numbers ($l$ and $l+1$) to the energy (\ref{energy}) is  (up to a constant $E_0$):
\begin{equation}
E_l=-\frac{\kappa_l}{g}, \label{energy_l}
\end{equation}

Now we go on to construct an effective spin-chain model, by replacing the spin-order index with the lattice site index in Eq.(\ref{psi_f_many_3}). In order to obtain the same energy functional as (\ref{energy_l}), the only way is to consider the following effective Hamiltonian
\begin{equation}
H_l=-\frac{J_l}{g} P_{00}(l,l+1),\label{Heff_f}
\end{equation}
with $P_{00}(l,l+1)$ the projection operator for neighboring sites ($l$ and $l+1$) forming a singlet, and $J_l$ follows
\begin{eqnarray}
J_l&=&2N! (\frac{\hbar^2}{m})^2 \int d{\bf x}  \Big|\frac{\partial \phi_F}{\partial x_{ij}}|_{x_{ij}=0} \Big|^2 \theta(\cdots<x_i=x_j<\cdots) ,   \nonumber\\     \label{J_l}
\end{eqnarray}
By expanding $P_{00}(l,l+1)$ in terms of the Pauli matrix and also noting that the total Hamiltonian is the summation of all neighboring-pair contributions, $H_{\rm eff}=\sum_l H_l$, finally we arrive at the effective spin-chain model for spin-1/2 fermions:
\begin{equation}
H_{\rm eff}=\sum_l \frac{J_l}{g} ({\bf s}_l\cdot{\bf s}_{l+1}-\frac{1}{4}).
\end{equation}
The result of anti-ferromagnetic correlation in above Hamiltonian is consistent with the Lieb-Mattis theorem saying that the ground state of spin 1/2 fermions is with the lowest total spin\cite{Lieb-Mattis}, as well as the indication from Bethe-ansatz solutions\cite{Guan_f}.

\bigskip

\subsubsection{spin-1/2 bosons}

For two-component (with pseudo-spin $\uparrow,\downarrow$) bosons, the interaction can occur in three channels:
\begin{equation}
U=\sum_{\sigma=\uparrow,\downarrow}g_{\sigma\sigma}\sum_{i<j} \delta(x_{i\sigma}-x_{j\sigma})+g_{\uparrow\downarrow}\sum_{i,j} \delta(x_{i\uparrow}-x_{j\downarrow}),
\end{equation}
These channels actually represent three spin-triplet channels due to the Bose statistics.

For large $g_{\sigma\sigma'}$, the many-body wave function can be written as certain superposition of the degenerate ground states (Eq.(\ref{Psi_f})) at all $1/g_{\sigma\sigma'}=0$:
\begin{eqnarray}
\Psi(\{x_i\};\{\mu_i\})=\Big|\phi_F(\{x_i\})\Big| \sum_{k=1}^{N_{\rm dg}} a_k \langle \{x_i\};\{\mu_i\} |\vec{\xi}_k \rangle,
\label{psi_b_many}
\end{eqnarray}
here the wave function is distinct from Eq.(\ref{psi_f_many}) by replacing $\phi_F$ with its absolute value. This replacement will significantly affect the expression of $\kappa_{\sigma\sigma'}$ as defined in the energy expansion at small $1/g_{\sigma\sigma'}$:
\begin{equation}
E=E_0-\frac{\kappa_{\uparrow\uparrow}}{g_{\uparrow\uparrow}}-\frac{\kappa_{\downarrow\downarrow}}{g_{\downarrow\downarrow}}-\frac{\kappa_{\uparrow\downarrow}}{g_{\uparrow\downarrow}}, \label{energy_b}
\end{equation}
with $\kappa_{\sigma\sigma'}$ given by:
\begin{equation}
\kappa_{\sigma\sigma'}= \frac{\partial E}{\partial (-1/g_{\sigma\sigma'})}=g_{\sigma\sigma'}^2 \frac{\partial E}{\partial g_{\sigma\sigma'}}.  \label{kappa_b}
\end{equation}
Following the similar procedure as in the last section, we obtain $\kappa_{\sigma\sigma'}$ as:
\begin{widetext}
\begin{eqnarray}
\kappa_{\uparrow\uparrow}&=&\frac{N(N-1)}{2} (\frac{\hbar^2}{m})^2 \int d{\bf x}  \Big|\frac{\partial \phi_F}{\partial x_{ij}}|_{x_{ij}=0} \Big|^2  \Big|\sum_{k=1}^{N_{\rm dg}} a_k \delta_{\mu_i,\uparrow}\delta_{\mu_j,\uparrow} \Big( \langle \{x_i\};\{\mu_i\} |\vec{\xi}_k \rangle|_{x_{ij}=0^+}+\langle \{x_i\};\{\mu_i\} |\vec{\xi}_k \rangle|_{x_{ij}=0^-} \Big) \Big|^2 ;\label{kappa_upup}  \\
\kappa_{\downarrow\downarrow}&=&\frac{N(N-1)}{2} (\frac{\hbar^2}{m})^2 \int d{\bf x}  \Big|\frac{\partial \phi_F}{\partial x_{ij}}|_{x_{ij}=0} \Big|^2  \Big|\sum_{k=1}^{N_{\rm dg}} a_k \delta_{\mu_i,\downarrow}\delta_{\mu_j,\downarrow} \Big( \langle \{x_i\};\{\mu_i\} |\vec{\xi}_k \rangle|_{x_{ij}=0^+}+\langle \{x_i\};\{\mu_i\} |\vec{\xi}_k \rangle|_{x_{ij}=0^-} \Big) \Big|^2 ;\label{kappa_dndn}  \\
\kappa_{\uparrow\downarrow}&=&N(N-1) (\frac{\hbar^2}{m})^2 \int d{\bf x}  \Big|\frac{\partial \phi_F}{\partial x_{ij}}|_{x_{ij}=0} \Big|^2  \Big|\sum_{k=1}^{N_{\rm dg}} a_k \delta_{\mu_i,\uparrow}\delta_{\mu_j,\downarrow} \Big( \langle \{x_i\};\{\mu_i\} |\vec{\xi}_k \rangle|_{x_{ij}=0^+}+\langle \{x_i\};\{\mu_i\} |\vec{\xi}_k \rangle|_{x_{ij}=0^-} \Big) \Big|^2 ;\label{kappa_updn}
\end{eqnarray}
\end{widetext}
Again we select out two coordinates, $x_i$ and $x_j$, staying in the neighboring orders, $l$-th and $l+1$-th, in the wave function, and denote their contribution to $\kappa_{\sigma\sigma'}$ as $\kappa_{\sigma\sigma';l}$. After simple algebra, we find that only the three triplet can contribute; explicitly, $|11\rangle_{l,l+1}, \ |1,-1\rangle_{l,l+1}$ and $|1,0\rangle_{l,l+1}$ respectively contribute to $\kappa_{\uparrow\uparrow},\ \kappa_{\downarrow\downarrow}$ and $\kappa_{\uparrow\downarrow}$. By rewriting the spin-ordered state as the form of Eq.(\ref{psi_f_many_3}), we can then obtain
\begin{eqnarray}
\kappa_{\uparrow\uparrow;l}&=&J_l \sum_n\Big|a_n^{11}\Big|^2,        \label{kappa_upup} \\
\kappa_{\uparrow\downarrow;l}&=&J_l \sum_n\Big|a_n^{10}\Big|^2,        \label{kappa_upup}\\
\kappa_{\downarrow\downarrow;l}&=&J_l \sum_n\Big|a_n^{1,-1}\Big|^2,        \label{kappa_upup}
\end{eqnarray}
here $J_l$ follows the same expression as Eq.(\ref{J_l}). Considering the energy functional (\ref{energy_b}), and replacing the spin-ordered index in (\ref{psi_f_many_3}) as the lattice site index, we can write down the effective spin chain model as
\begin{eqnarray}
H_{\rm eff}&=&-\sum_l J_l \left( \frac{1}{g_{\uparrow\uparrow}} P_{11}(l,l+1) +  \frac{1}{g_{\downarrow\downarrow}} P_{1,-1}(l,l+1) \right. \nonumber\\
&&\ \ \ \ \ \  \left.+ \frac{1}{g_{\uparrow\downarrow}} P_{10}(l,l+1) \right), \label{Heff_b}
\end{eqnarray}
with $P_{SM}(l,l+1)$ the projection operator for neighboring sites $l$ and $l+1$ forming a triplet with $S=1,\ M=0,\pm 1$.
For the case of $g_{\uparrow\uparrow}=g_{\downarrow\downarrow}\equiv g$,  Eq.(\ref{Heff_b}) can be reduced to the XXZ Heisenberg model:
\begin{eqnarray}
H_{\rm eff}&=&\sum_l J_l \left(  -\frac{1}{g_{\uparrow\downarrow}} (s_l^xs_{l+1}^x+s_l^y s_{l+1}^y) +( \frac{1}{g_{\uparrow\downarrow}} - \frac{2}{g} ) s_l^z s_{l+1}^z \right. \nonumber\\
&&\ \ \ \ \ \left. - ( \frac{1}{4g_{\uparrow\downarrow}} + \frac{1}{2g} ) \right) . \label{model1_boson}
\end{eqnarray}
For the case of  $SU(2)$ interaction with spin-independent interaction strength $g_{\uparrow\uparrow}=g_{\downarrow\downarrow}=g_{\uparrow\downarrow}\equiv g$, Eq.(\ref{Heff_b}) is reduced to the isotropic Ferromagnetic Heisenberg model:
\begin{equation}
H_{eff}=-\sum_l\frac{J_l}{g} \left({\bf s}_l\cdot {\bf s}_{l+1} +\frac{3}{4}\right)   . \label{Heff_manybody_b}
\end{equation}
This is consistent with general theorems showing that the ground state of iso-spin 1/2 bosons is ferromagnetic\cite{FM1,FM2,FM3} and with the result from Bethe-ansatz solutions\cite{Guan_b}. We note that the effective models (Eqs.\ref{model1_boson},\ref{Heff_manybody_b}) preserve the full symmetry of original Hamiltonian, as they take the same structure as that of the interaction potentials. These models are related to those in Ref.\cite{Zinner3, Levinsen2} by a unitary transformation with operator $u=\prod_{i=1}^{[N/2]}\sigma_i^z$\cite{note_boson_model}.

Comparing spin-1/2 bosons with fermions, we can see that the distinct correlations (FM for bosons and AFM for fermions) in Heisenberg Hamiltonians are intrinsically resulted from the different symmetries allowed for the system. For the fermions, the anti-symmetric feature of the wave function requires that the particles scatter in the singlet channel, while for bosons the symmetric wave function requires the triplet channels.
These scattering channels uniquely determine the structure of the effective models (see Eqs.(\ref{Heff_f},\ref{Heff_b})) in terms of the projector operators $P_{SM}$. %, while %$J_l$ representing the amplitude of local scattering of two particles.
This structure manifests the intrinsic relation between the statistics, the interaction channels, the effective spin-chain Hamiltonian and the nature of the ground state. In the following, we will show that it can be straightforwardly generalized to an arbitrary high-spin case.

%A similar separation of the density and spin excitations is well known in the case of strongly interacting one-dimensional fermions, where it was first derived [13] from the exact solution of the infinite-U Hubbard model. The sign of the exchange constant J is determined by the requirement to either symmetrize or antisymmetrize the wave function with respect to the permutation; the coupling is antiferromagnetic for fermions, J <0, and ferromagnetic for bosons, J >0. On the other hand, the magnitude of the exchange constant J is determined by the amplitude of the forward scattering of two neighboring particles, regardless of their statistics. Thus we find the same value of J as in the case of fermions with strong short-range repulsion,

%\bigskip
\subsection{Effective spin-chain model for a general spin}

For a general high-spin system with multi-channel scatterings, the interaction can be written as :
\begin{eqnarray}
U&=&\sum_{S=0}^{2f}\sum_{M=-S}^{S} g_{SM}  \sum_{i<j} \delta(x_i-x_j) P_{SM}(i,j). \label{U2}
\end{eqnarray}
Here $g_{SM}$ is the coupling constant for two particles scattering with total spin $S$ and total magnetization $M$; $P_{SM}(i,j)$ is the projection operator for particles $i,j$ scattering in the $\{SM\}$ sector.

For large $g_{SM}$, the many-body wave function of bosons (B) and fermions (F) can be respectively written as:
\begin{eqnarray}
\Psi_B(\{x_i\};\{\mu_i\})&=&\Big|\phi_F(\{x_i\})\Big| \sum_{k} a_k \langle \{x_i\};\{\mu_i\} |\vec{\xi}_k \rangle,
\label{psi_b_many_2} \\
\Psi_F(\{x_i\};\{\mu_i\})&=&\phi_F(\{x_i\}) \sum_{k} a_k \langle \{x_i\};\{\mu_i\} |\vec{\xi}_k \rangle.
\label{psi_f_many_2}
\end{eqnarray}
The energy expansion at small $1/g_{SM}$ can be expanded as:
\begin{equation}
E=E_0-\sum_{SM} \frac{\kappa_{SM}}{g_{SM}}, \label{energy_bf}
\end{equation}
with $\kappa_{SM}$ given by:
\begin{equation}
\kappa_{SM}= \frac{\partial E}{\partial (-1/g_{SM})}=g_{SM}^2 \frac{\partial E}{\partial g_{SM}}.  \label{kappa_bf}
\end{equation}
The boundary condition around $x_{ij}\equiv x_i-x_j=0$ gives  %we have
\begin{eqnarray}
\frac{\hbar^2}{m} \frac{\partial (P_{SM}\Psi)}{\partial x_{ij}}|_{x_{ij}=0^-}^{x_{ij}=0^+} &=&g_{SM} (P_{SM}\Psi)|_{x_{ij}=0},
\end{eqnarray}
with $\Psi$ the many-body wave function for bosons or fermions. Based on this boundary condition, we can expand (\ref{kappa_bf}) as
\begin{widetext}
\begin{eqnarray}
\kappa_{SM}&=&\frac{N(N-1)}{2} (\frac{\hbar^2}{m})^2 \int d{\bf x}  \Big|\frac{\partial \phi_F}{\partial x_{ij}}|_{x_{ij}=0} \Big|^2  \Big|\sum_{k} a_k P_{SM} \Big( \langle \{x_i\};\{\mu_i\} |\vec{\xi}_k \rangle|_{x_{ij}=0^+}-\langle \{x_i\};\{\mu_i\} |\vec{\xi}_k \rangle|_{x_{ij}=0^-} \Big) \Big|^2 ;\label{kappa_f}
\end{eqnarray}
for fermions, and
\begin{eqnarray}
\kappa_{SM}&=&\frac{N(N-1)}{2} (\frac{\hbar^2}{m})^2 \int d{\bf x}  \Big|\frac{\partial \phi_F}{\partial x_{ij}}|_{x_{ij}=0} \Big|^2  \Big|\sum_{k}a_k P_{SM} \Big( \langle \{x_i\};\{\mu_i\} |\vec{\xi}_k \rangle|_{x_{ij}=0^+}+\langle \{x_i\};\{\mu_i\} |\vec{\xi}_k \rangle|_{x_{ij}=0^-} \Big) \Big|^2 ;\label{kappa_b}
\end{eqnarray}
for bosons.
\end{widetext}
Assume two coordinates, $x_i$ and $x_j$, staying in the neighboring orders, $l$-th and $l+1$-th, in the wave function, and denote their contribution to $\kappa_{SM}$ as $\kappa_{SM;l}$. Due to the sign difference in the expressions of $\kappa_{SM}$ for bosons and fermions, we find that only when the spin wave function in the spin-ordered states is symmetric for bosons and antisymmetric for fermions, can it contribute to  $\kappa_{SM;l}$. This can also select out the spin channels in which the particles can scatter with each other.

By rewriting the spin-ordered state as the form of Eq.(\ref{psi_f_many_3}), we can obtain
\begin{eqnarray}
\kappa_{SM;l}&=&J_l \sum_n\Big|a_n^{SM}\Big|^2,        \label{kappa_SM}
\end{eqnarray}
here $J_l$ follows the same expression as Eq.(\ref{J_l}). Considering the energy functional (\ref{energy_bf}), and replacing the spin-ordered index in (\ref{psi_f_many_3}) as the lattice site index, we can write down the effective spin chain model as
\begin{eqnarray}
H_{\rm eff}&=&-\sum_l J_l \sum_{SM} \frac{1}{g_{SM}} P_{SM}(l,l+1) , \label{Heff_bf}
\end{eqnarray}
with $P_{SM}(l,l+1)$ the projection operator for neighboring sites $l$ and $l+1$ forming a state with total spin $S$ and total magnetization $M$. Above results can be easily applied to the spin-1/2 case of fermions ($S=M=0$) and bosons ($S=1,M=0,\pm 1$), and accordingly the AFM or FM Heisenberg models can be obtained respectively.

To this end, we have obtained a general form of effective spin-chain model, Eq.(\ref{Heff_bf}), to describe strongly coupling atoms in 1D trapped system. It can be applied for any statistics (Bose or Fermi), an arbitrary spin, and any spin-independent confinement potential.
%Above results can be easily applied to the spin-1/2 case. For spin-1/2 fermions, only the singlet channel is allowed by the symmetry, $S=M=0$, and the resulted effective model for the two sites is simply proportional to the singlet spin projection operator of these two sites. For spin-1/2 bosons, by symmetry the particles are allowed to scatter in three triplet channels, and the resulted spin-chain model are contributed by three channels. From this, one can see clearly how the statistics affects the form of the spin-chain model.
A key property of this effective model is that locally for each site $l$ the Hamiltonian can be well separated into two parts: the coupling parameter $J_l$ which only relies on the local scattering amplitude, and the nearest-neighbor spin-projection which follows the same structure as the interaction model.
% does not depend on the spin at all, but only rely on the local density at site (or order) index $l$; while the relevance of the spin part is all incorporated by the projection terms.
This is closely related to the spin-charge separation in the wave function in the hard-core limit, as the form of Eqs.(\ref{psi_b_many_2},\ref{psi_f_many_2}). More explicitly, the charge part ($\phi_F$) determines the local scattering amplitude $J_l$, while the spin-ordered part together with the statistics determine the spin-relevant terms in Eq.(\ref{Heff_bf}).

% terms in the Jeff is solely determined by the charge part ($\phi_F$), while the rest terms are given by the spin-ordered part and reflect the statistics.

\section{Strongly coupling spin-1 bosons and the effect of spin-orbit coupling}

In this section, we will apply the effective spin-chain model (\ref{Heff_bf}) to the spin-1 bosons. At low fields, the interaction of spin-1 bosons has $SU(2)$ symmetry and can be classified into two channels:
\begin{eqnarray}
U&=& \sum_{i<j} \delta(x_i-x_j) \left( g_0P_{S=0}(i,j) +  g_2P_{S=2}(i,j) \right) . \label{U_spin1}
\end{eqnarray}
here $g_0$ and $g_2$ are respectively the coupling constant in the total spin $S=0$ and $S=2$ channels\cite{spin1_H_1,spin1_H_2}, and $P_S$ is the projection operator to the total spin $S$-channel. Depending on the relative strength of $g_0$ and $g_2$, the cold atoms system can have AFM (or polar/nematic) ground state ($^{23}$Na with $0<g_0<g_2$), or FM ground state ($^{87}$Rb with $0<g_2<g_0$).

Based on the general formula (\ref{Heff_bf}), we can obtain the effective spin-chain model for spin-1 bosons in the strong coupling limit:
\begin{eqnarray}
H_{\rm eff}&=&-\sum_l J_l \left( \frac{1}{g_{0}} P_{0}(l,l+1) + \frac{1}{g_{2}} P_{2}(l,l+1) \right) , \label{Heff_spin1}\nonumber\\
\end{eqnarray}
which can be reduced to the form:
\begin{eqnarray}
H_{\rm eff}&=&-\frac{C_N}{g_0}  \sum_l j_l \left( b_2 ({\bf s}_l \cdot {\bf s}_{l+1})^2 +  b_1 {\bf s}_l\cdot{\bf s}_{l+1} + b_0 \right) , \label{Heff_spin1_2}\nonumber\\
\end{eqnarray}
with ${\bf s}_l$ the spin operator at site $l$. In writing above equation, $J_l$ has been decomposed as the product of $C_N$ and $j_l$. $C_N $ is a quantity that only depends on the total particle number and the underlying trapping potential, which is proportional to the contact of the system, while the site-dependence of $J_l$ is simply included in $j_l$. Since $j_l$ does not depend on the spin, here we use a good approximation derived for the spin-1/2 fermions case in  a harmonic trap\cite{Levinsen}:
\begin{equation}
j_l=\frac{-(l-N/2)^2+N^2/4}{N(N-1)/2}\ \ \ \ \ (l=1,...,N-1),
\end{equation}
$b_2,\ b_1,\ b_0$ in Eq.\ref{Heff_spin1_2} are dimensionless parameters depending on the ratio of interaction strengths in different channels:
\begin{eqnarray}
b_2&=&\frac{1}{3}(1+\frac{g_0}{2g_2}); \\
b_1&=&\frac{g_0}{2g_2}; \\
b_0&=&\frac{1}{3}(\frac{g_0}{g_2}-1).
\end{eqnarray}
Note that the structure of the Hamiltonian (\ref{Heff_spin1_2}) is the same as that of the bilinear-biquadratic isotropic quantum $S=1$ chain and that in the Mott phase of spin-1 bosons in optical lattices \cite{spin1_1,spin1_2,spin1_3}. The only difference is that here the coupling constants of neighbor spins are site-dependence, due to the background trap potential and the resultant inhomogeneous particle (charge) density.

When add the spin-orbit coupling(SOC) to the system, which corresponds to a rotating field in the coordinate space\cite{Ian_1, Ian_2, Shuai, Jing, Martin,Engels}:
\begin{eqnarray}
H_{\rm SOC}&=&-\Omega_R\int dx \sum_m \left( e^{iqx} \psi^{\dag}_{m}(x)\psi_{m+1}(x) +h.c.\right),\nonumber\\
\end{eqnarray}
the spin texture and the spin-spin correlation according to $H_{\rm eff}$ (Eq.\ref{Heff_spin1_2}) can be greatly modified.
% tuned by the strength of the SOC and the rotating angle of the field at each site.
In the basis of spin-ordered states, the SOC Hamiltonian can be greatly reduced similarly to the spin-1/2 fermion case\cite{Cui3}:
\begin{eqnarray}
H_{\rm SOC}
&=&-\frac{\Omega}{\sqrt{2}}\sum_{l=1}^{N}  \left( e^{il\phi } s_-^l +h.c.\right).  \label{hsoc}
\end{eqnarray}
Here $\Omega=\Omega_R e^{-q^2\bar{\sigma}^2/4}$ and $\phi=qd$\cite{Cui3}, with $\bar{\sigma}$ the averaged $\sigma_l$ defined in Eq.\ref{ni} and $d$ the mean particle distance $d=\overline{\bar{x}_{l+1}-\bar{x}_{l}}$. We make such approximation by noting that $\sigma_l$ and $\bar{x}_{l+1}-\bar{x}_{l}$ differ little between different $i$, as shown in Table I. Moreover, here as we are limited at small $\Omega_R(\ll \omega_T)$ and moderate $\phi$, we have neglected the contribution from higher harmonic levels and thus the extra spin-exchange terms as pointed out in Ref.\cite{Blume2}.

\begin{figure}[t]
\includegraphics[height=6cm,width=8.8cm]{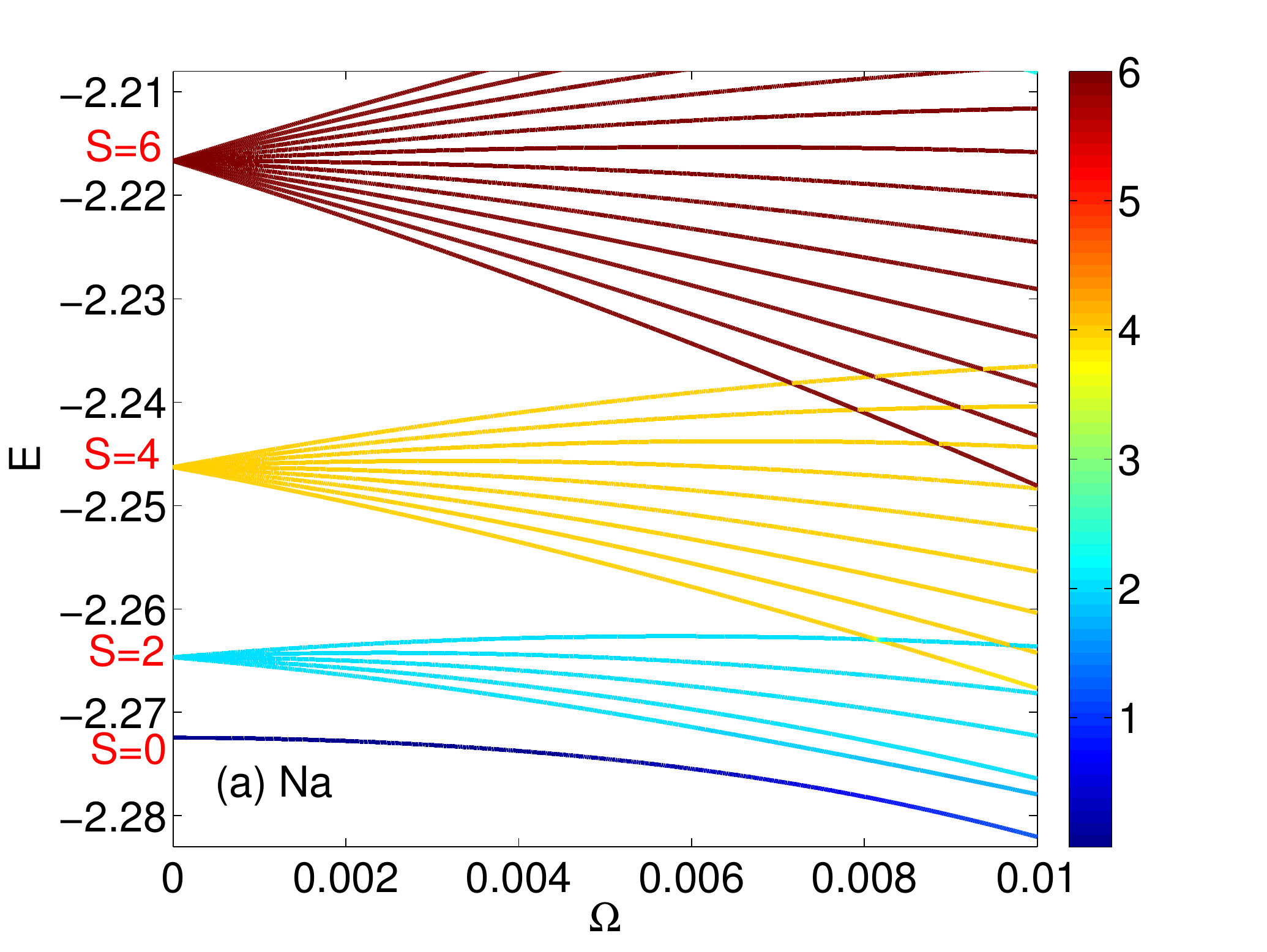}
\includegraphics[height=6cm,width=8.8cm]{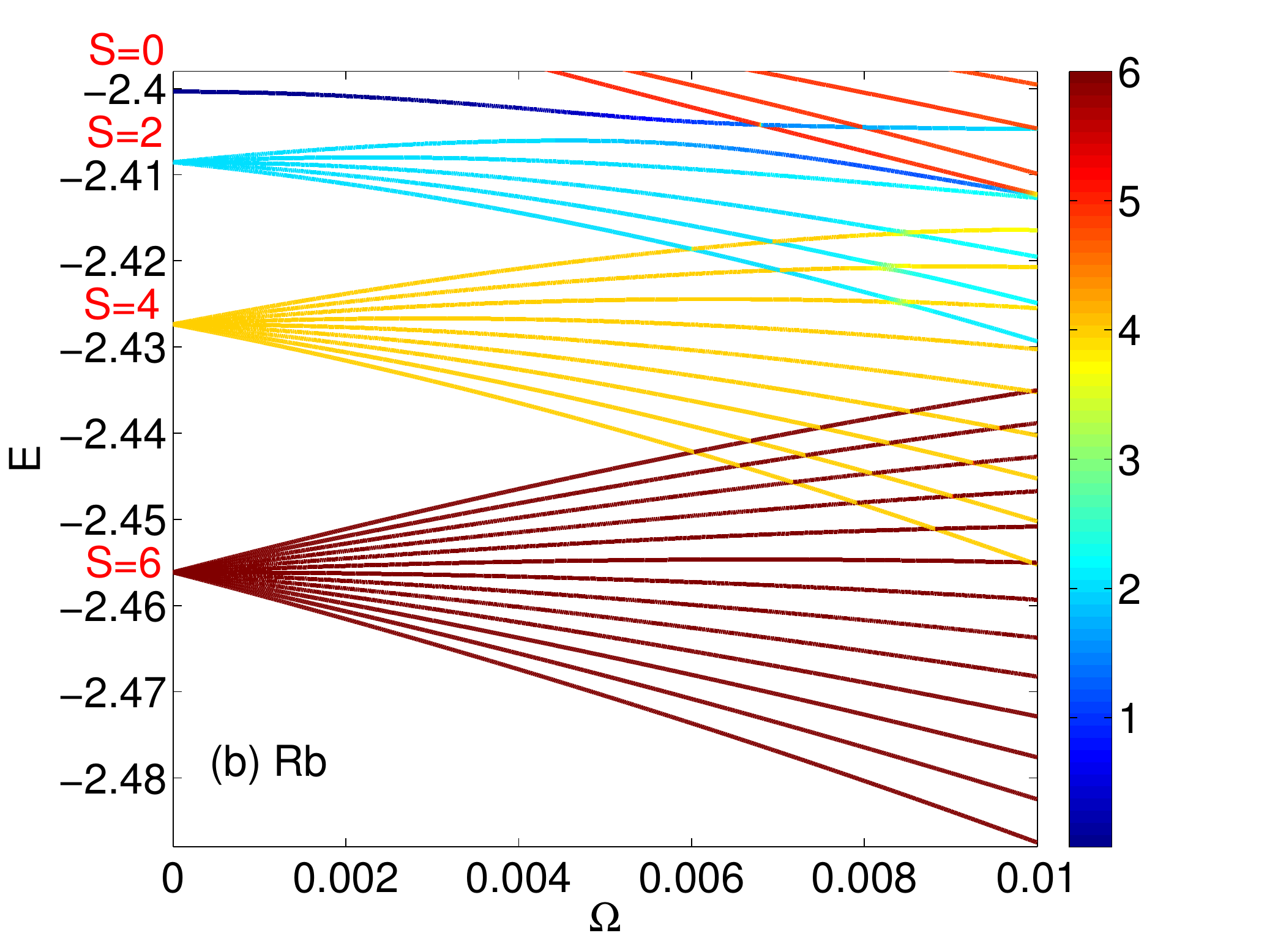}
\caption{(Color online). Energy spectrum as a function of spin-orbit coupling strength $\Omega$ for $^{23}$Na (a) and$^{87}$Rb (b) atoms with total number $N=6$. $\phi$ is chosen to be $\pi/4$. All energies are in the unit of $C_N/g_0$. Different colors represent different spin values determined by $S(S+1)\equiv \langle {\bf S}^2 \rangle$.  } \label{fig2}
\end{figure}

In Fig.2, we show the energy spectrum as a function of $\Omega$ for both Na ($g_0<g_2$) and Rb ($g_0>g_2$) systems with particle number $N=6$ and rotation angle $\phi=\pi/4$. In the absence of SOC ($\Omega=0$), the total spin is conserved. For Na, the ground state is a singlet with $S=0$, and as $S$ increases the energy also increases. For Rb, the situation is reversed: the ground state is highly degenerate with the largest total spin $S=N=6$, and the energy increases as $S$ decreases. For both Na and Rb, we can see that the states are degenerate for different magnetization $M$ with the same $S$. However, this is no longer true when turn on even a tiny strength of SOC ($\Omega>0$). In this case, the original degeneracies are completely lifted, as shown by Fig.2(a,b), because the SOC field in (\ref{hsoc}) does not commutate with the total spin.

\begin{figure}[t]
\includegraphics[height=5.5cm,width=8.5cm]{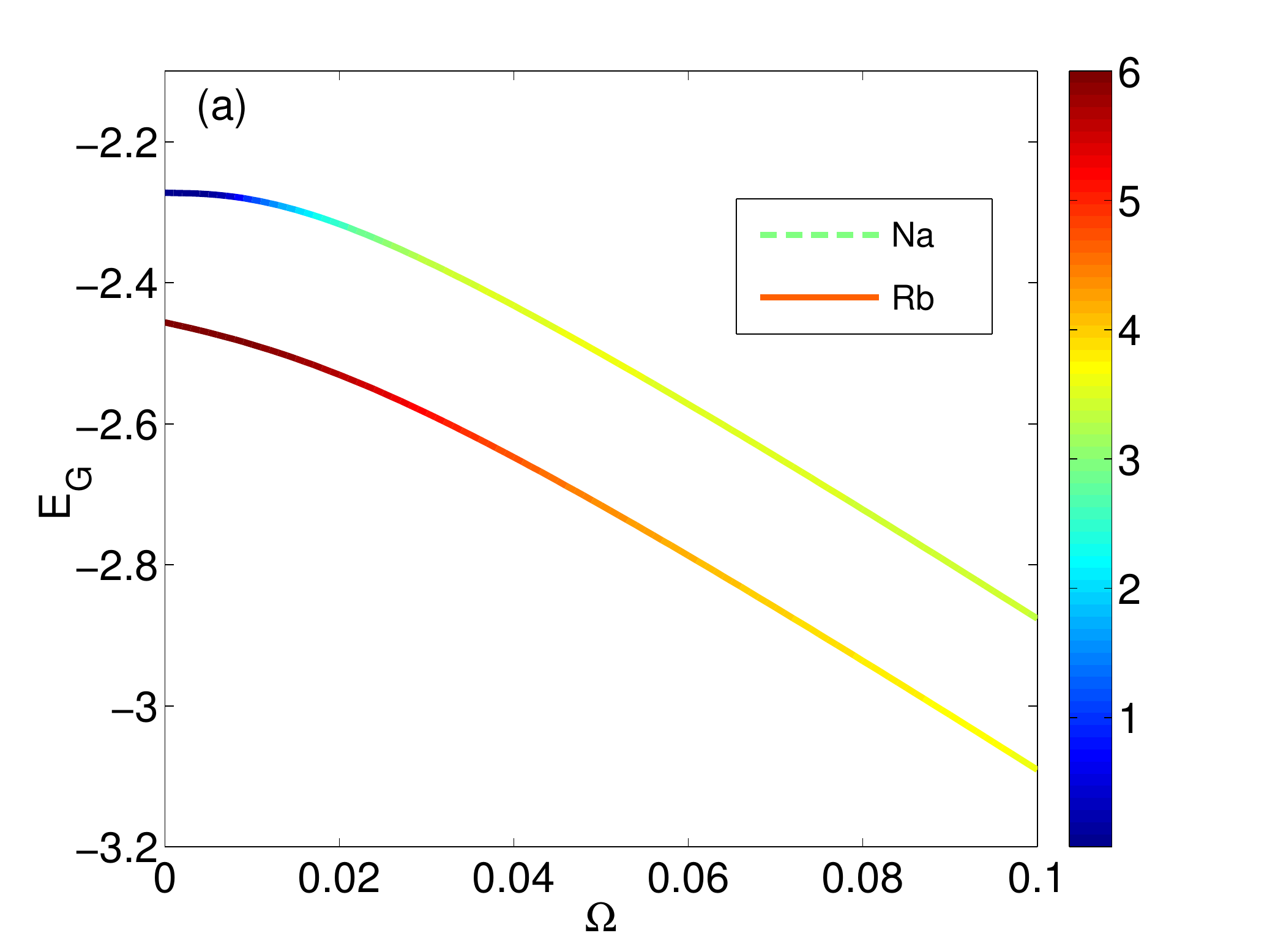}
\includegraphics[height=5.5cm,width=7.5cm]{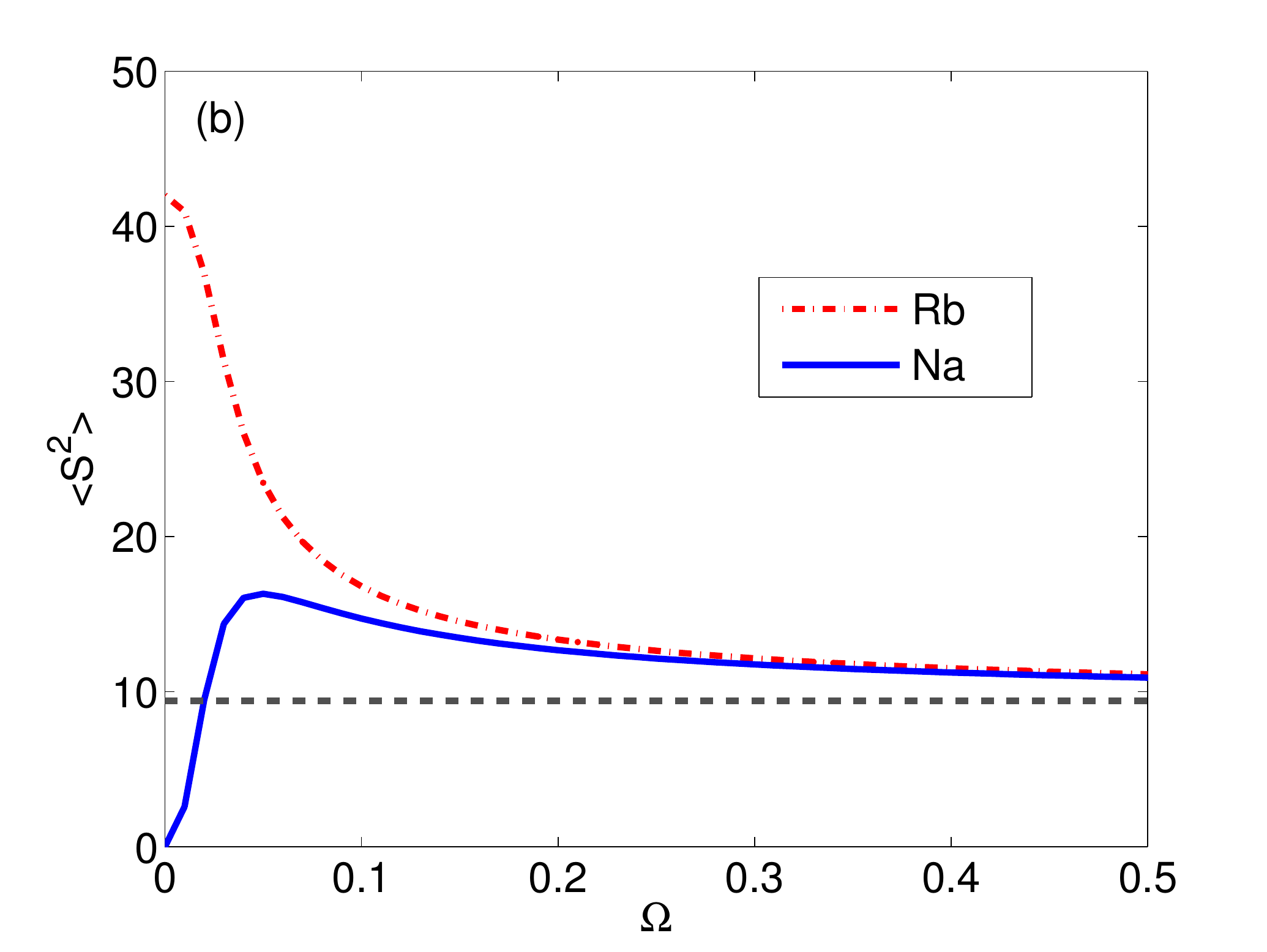}
\caption{(Color online). The ground state energy $E_G$(a) and total spin $\langle {\bf S}^2\rangle$ (b) as a function of $\Omega$ for both Na and Rb. All parameters are the same as in Fig.2. Gray dashed line in (b) shows the universal value in the SOC dominated regime.} \label{fig3}
\end{figure}

In Fig.3, we show the ground state energy $E_G$ and $\langle {\bf S}^2\rangle$ as a function of  $\Omega$. We see that $E_G$ of both Na and Rb decrease as increasing $\Omega$, while the total spin tend to approach the same value at $\Omega\ge 0.4C_N/g_0$. In this regime, the AFM or FM correlation in the spin chain model is fully destroyed by the SOC field, and the systems of both Na and Rb are fully governed by the local rotating field in $H_{SOC}$(Eq.(\ref{hsoc})). Thus increasing $\Omega$, the system will undergo a crossover from the interaction-dominated to the SOC-dominated regime.

\begin{figure}[h]
\includegraphics[height=6.5cm,width=9.5cm]{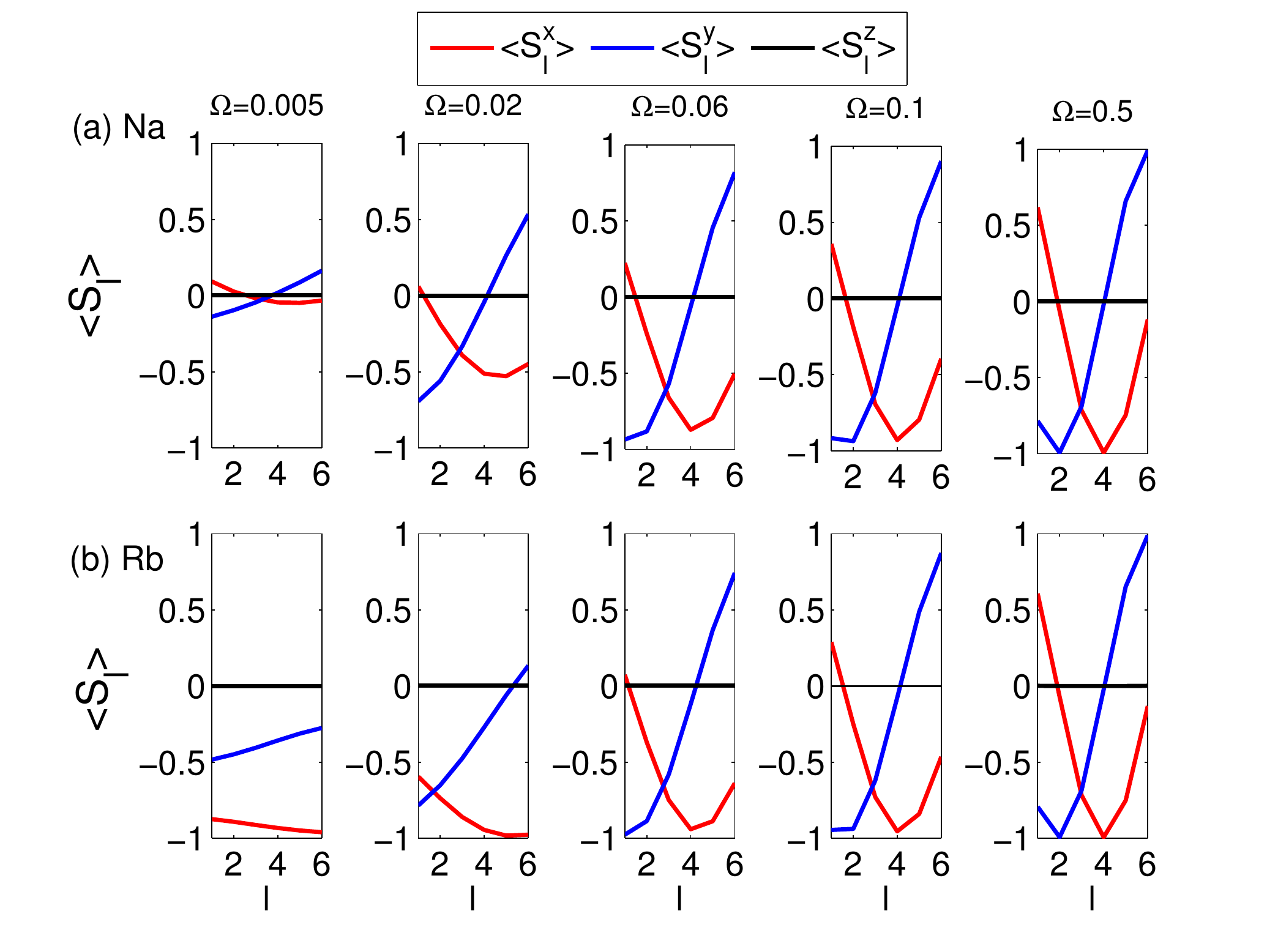}
\caption{(Color online). Local spin $\langle {\bf s}_l\rangle$ in the ground state at several values of $\Omega$ for Na (upper panel) and Rb (lower panel) atoms. All parameters are the same as in Fig.2.  } \label{fig4}
\end{figure}

\begin{figure}[t]
\includegraphics[height=5.5cm,width=8cm]{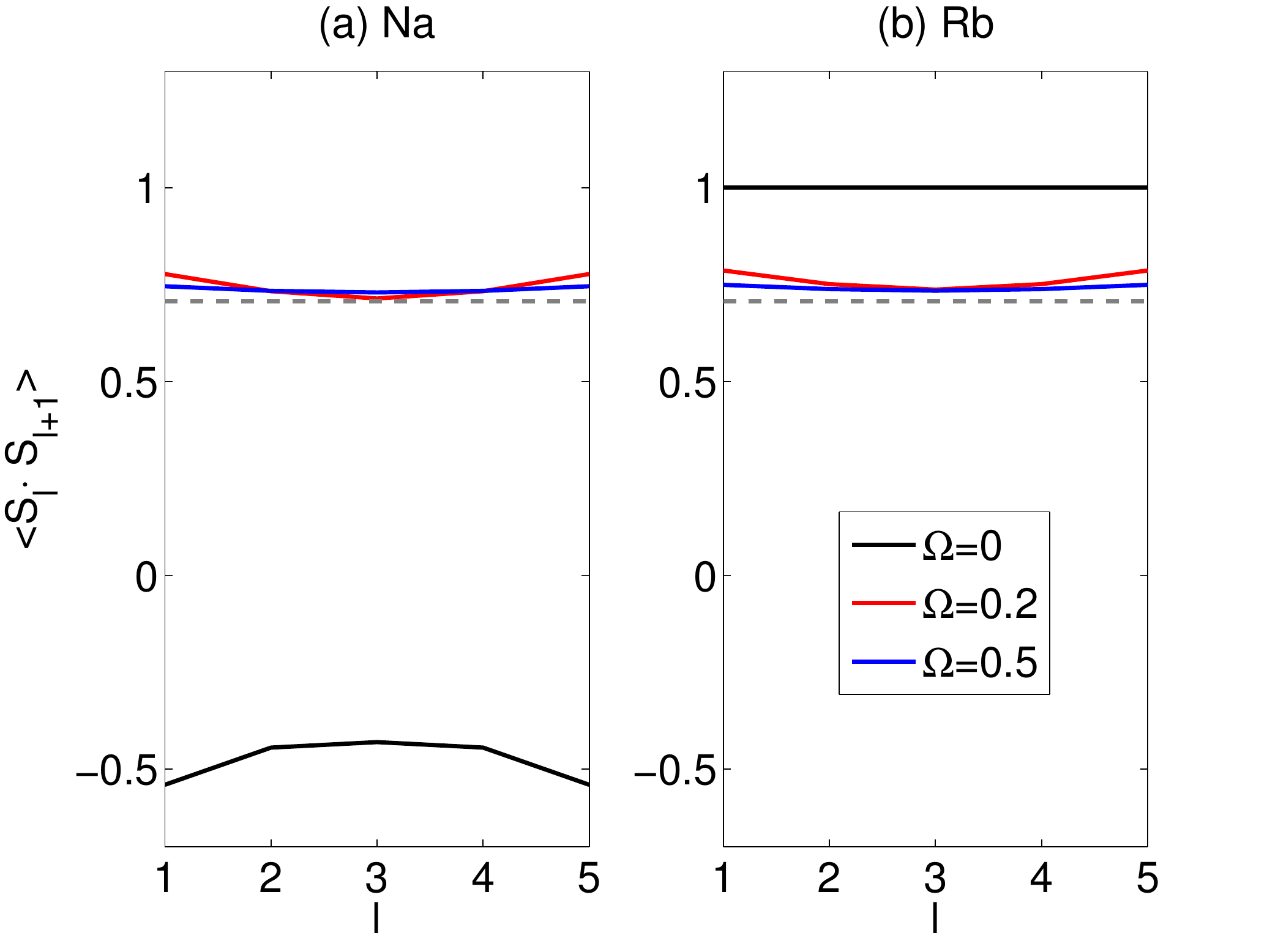}
\caption{(Color online). Spin-spin correlation $\langle {\bf s}_l \cdot {\bf s}_{l+1}\rangle$ in the ground state at several values of $\Omega$ for Na (a) and Rb (b) atoms. All parameters are the same as in Fig.2. Gray dashed lines show the universal value in the SOC dominated regime. } \label{fig5}
\end{figure}

\begin{figure}[t]
\includegraphics[height=6cm,width=8.8cm]{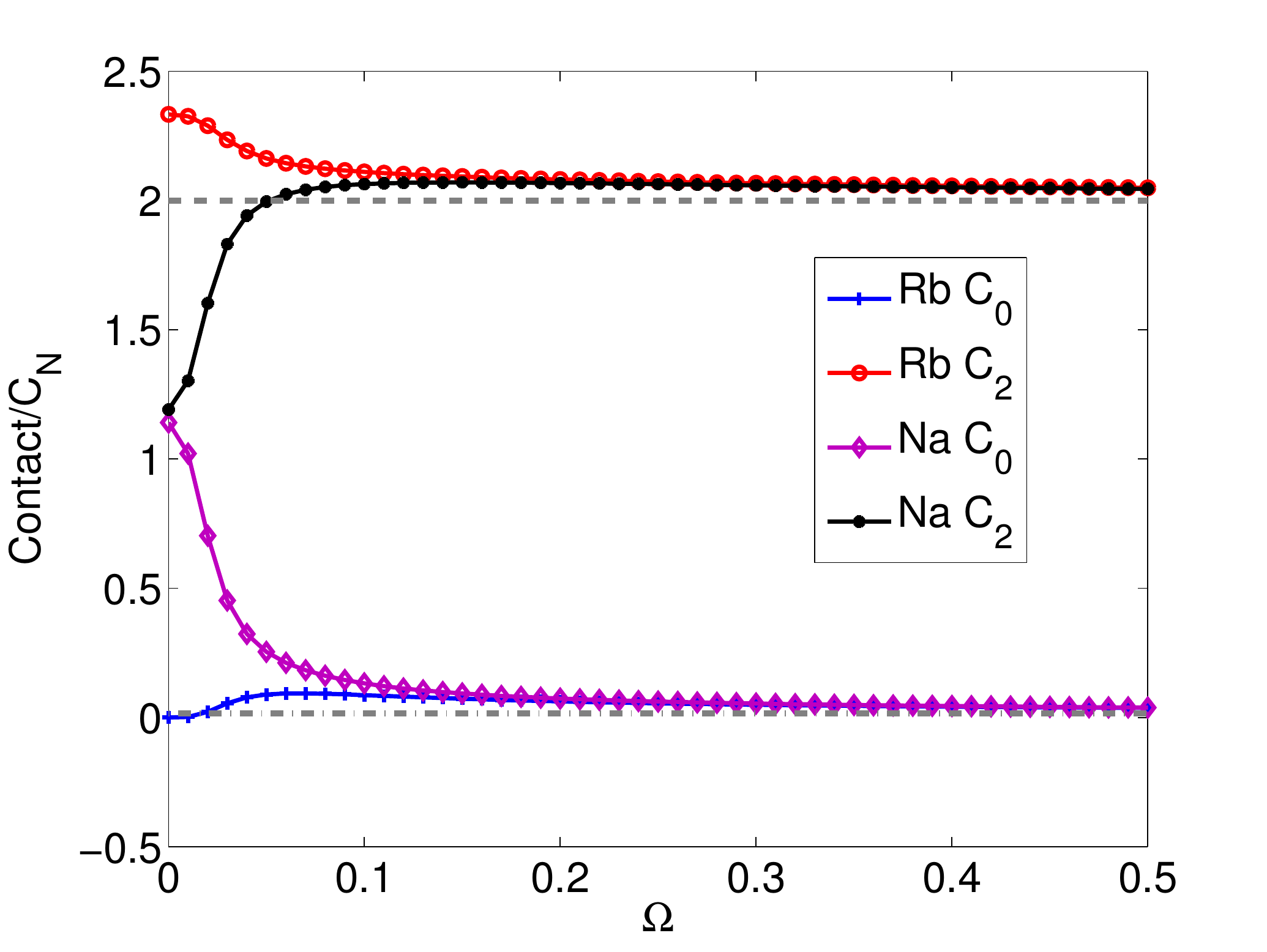}
\caption{(Color online). Contact $C_{\alpha}$ ($\alpha=0,2$, in the unit of $C_N$)
for the ground state as a function of $\Omega$ for Na and Rb atoms. All parameters are the same as in Fig.2. Gray dashed lines show the universal values in the SOC dominated regime. } \label{fig6}
\end{figure}

\begin{figure}[h]
\includegraphics[height=6cm,width=8.8cm]{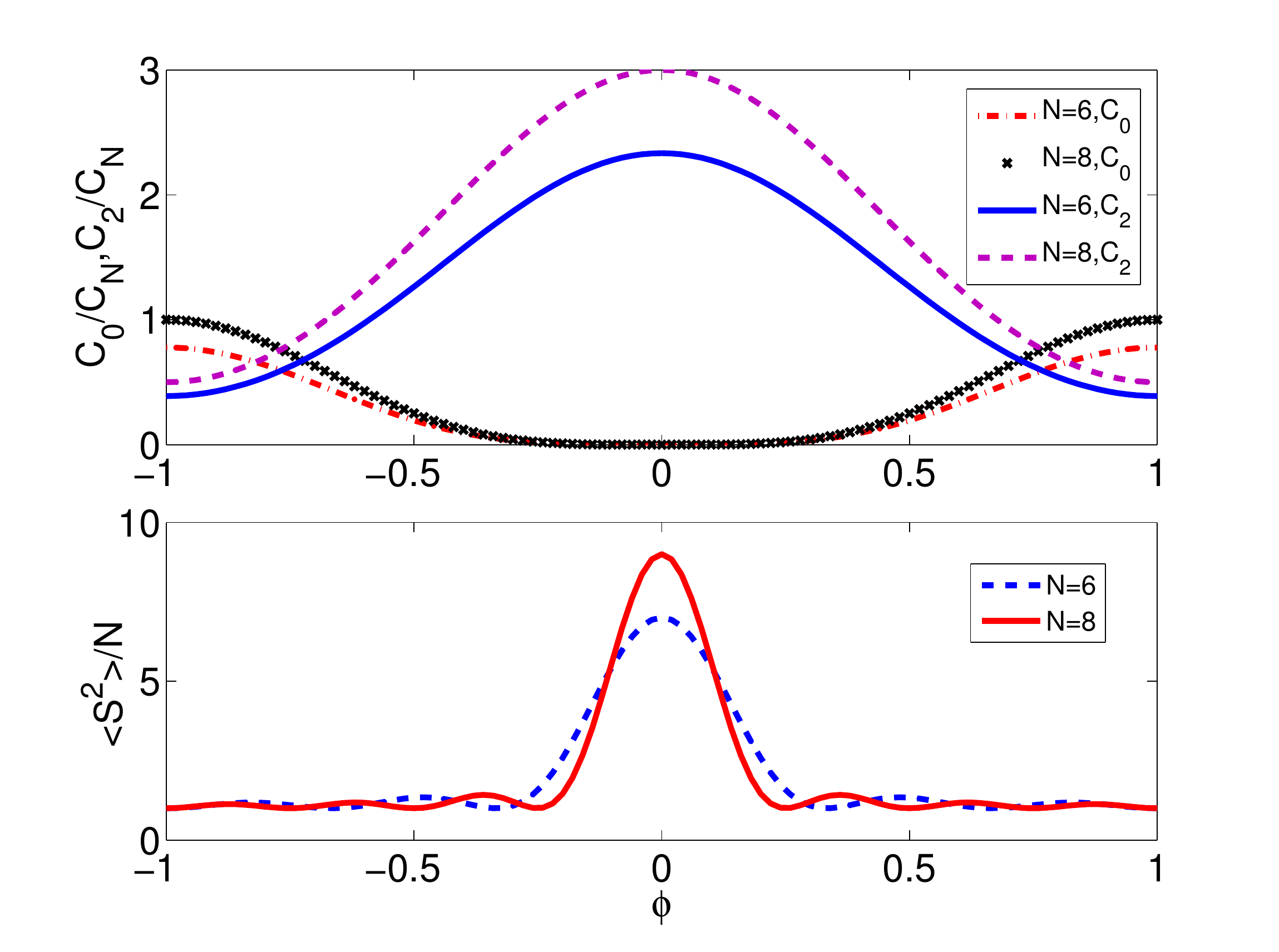}
\caption{(Color online). The contacts $C_{0},  C_2$ (in the unit of $C_N$, upper panel) and the total spin $\langle S^2\rangle/N$ as functions of rotation angle $\phi$ (in the unit of $\pi$) for the ground states of $N=6,8$ particles in the SOC-dominated regime.  } \label{fig7}
\end{figure}

One can also see the evidence of such crossover by examining the spin structure and the contact of the system. Fig.4, Fig.5 and Fig.6 respectively show the spin texture $\langle {\bf s}_l\rangle$, nearest-neighbor correlation $\langle {\bf s}_l \cdot {\bf s}_{l+1}\rangle$, and the contacts $C_{\alpha}=\partial E_G/\partial g_{\alpha}^{-1}$ ($\alpha=0,2$) for the ground states of Na and Rb at several values of $\Omega$. At zero or small $\Omega$, we can see clearly the signature of AFM or FM correlations: for Na,  $\langle {\bf s}_l\rangle\sim 0$ for all $i$ characterizing the polar/nematic ground state; for Rb, we have $|\langle {\bf s}_l\rangle|\sim 1$ for all $i$, $\langle {\bf s}_l \cdot {\bf s}_{l+1}\rangle=1$ for any nearest-neighbor pair, and $C_0=0$ all characterizing a FM state. As increasing $\Omega$, both Na and Rb will develop a large spin spiral with amplitude of the order of unity (see Fig.4). Consequently, $\langle {\bf s}_l \cdot {\bf s}_{l+1}\rangle$ and  $C_{\alpha}$ also approach universal values for both Na and Rb which do not change as increasing $\Omega$ further (see Fig.5 and Fig.6). When at the special point of infinite coupling, any infinitesimal $\Omega$ will induce a large and universal spin-spiral, as identified previously in the spin-1/2 fermions case \cite{Cui3}.

% on the actual strength of , which do large spin spiral with $\langle {\bf S}_l\rangle=(\cos(l\phi),\sin(l\phi),0)$, correspondingly a universal value of $\langle {\bf S}_l \cdot {\bf S}_{l+1}\rangle=\cos\phi(=\sqrt{2}/2 \ {\rm for}\ \phi=\pi/4)$ is reached. When at the infinite coupling, any infinitesimal $\Omega$ will induce a large and universal spin-spiral, as identified in the spin-1/2 fermions previously \cite{Cui3}. In the SOC-dominated regime, the two Contacts $C_0$ and $C_2$ also reach universal values for both Na and Rb, regardless of the strength of SOC.

In the SOC-dominated regime, each site $l$ in the spin-chain model (\ref{Heff_spin1_2}) is decoupled from others and the ground state can be straightforwardly obtained as
\begin{equation}
|\Psi_G\rangle=\prod_l \frac{1}{2} \left( e^{il\phi} c_{l,1}^{\dag} - \sqrt{2} c_{l,0}^{\dag} + e^{-il\phi} c_{l,-1}^{\dag} \right) |Vac\rangle,  \label{psi_soc}
\end{equation}
here $c_{l,m}^{\dag}$ is the creation operator of a single atom with magnetic number $m$ at site $l$. Given the wave function (\ref{psi_soc}), we can get the following universal physical quantities:
\begin{eqnarray}
&&\langle {\bf s}_l\rangle=(\cos(l\phi),\sin(l\phi),0);  \label{s_l}\\
&&\langle {\bf s}_l \cdot {\bf s}_{l+1}\rangle=\cos\phi; \label{ss}\\
&&\langle {\bf S}^2 \rangle=N+\sin^2(N\phi/2)/\sin^2(\phi/2) ; \label{s2}\\
&&C_0/C_N= (\frac{1}{12}-\frac16\cos\phi+\frac{1}{12}\cos^2\phi)\sum_l j_l ;\\
&&C_2/C_N =(\frac{13}{24}+\frac{5}{12}\cos\phi+\frac{1}{24}\cos^2\phi)\sum_l j_l ;
\end{eqnarray}
From Fig.3-6, we can see that these universal values fit the numerical results well in the SOC-dominated regime (with fixed $\phi$ and $N$). In Fig.7, we further show how these universal values of the contacts and the total spin depend on the rotation angle $\phi$ for six and eight particles.

\section{Summary}

In summary, we have presented a general form of effective spin-chain model for strongly-interacting 1D trapped systems, which is applicable for an arbitrary spin, any statistics (Bose or Fermi) and any spin-independent confinement potentials.  Importantly, this general model, as shown in Eq.(\ref{Heff_bf}), contains two essential ingredients. One is the local coupling parameter due to the inhomogeneity of charge density in the trapped system, which is irrelevant to spins. The other is the nearest-neighbor spin-projection, which generically follows the same structure as the interaction models and uniquely determines the spin-spin correlation and magnetic property of the system. Such effective model reflects the intrinsic relation between the statistics, scattering channels, and the nature of the ground state. It can serve as a useful and efficient tool to study the equilibrium and non-equilibrium properties of strongly interacting 1D trapped systems.

Take the spin-1 bosons for example, we further show how the addition of spin-orbit coupling(SOC) can destroy the original magnetic property according to the effective spin-chain model. Increasing the SOC strength, the system undergoes a crossover from the interaction-dominated to SOC-dominated regime, and eventually a ground state with universal spin structure and contacts will be achieved. Here the spin-chain model allows the study of the interplay between strong interaction, high-spin, and spin-orbit coupling in a convenient and physically transparent manner.

\section{Acknowledgement}
We would like to thank Nikolaj T. Zinner for helpful feedback on the manuscript. This work is supported by the National Natural Science Foundation of China under grant No.11374177, and the programs of Chinese Academy of Sciences.

\clearpage


\begin{thebibliography}{99}
\bibitem{note_density} The spreading property will crucially depend on the type of underlying trapping potential. Other potentials will have different properties\cite{Zinner3, Zinner4}.
\end{thebibliography}
\end{document}